\documentclass[aip,pop,reprint]{revtex4-1}

\usepackage{graphicx}	% Including figure files
\usepackage{amsmath}	% Advanced maths commands
\usepackage{hyperref}
\hypersetup{
    colorlinks=true,
    linkcolor=blue,
    urlcolor=blue,
    citecolor=blue,
    }

\begin{document}

% Use the \preprint command to place your local institutional report number 
% on the title page in preprint mode.
% Multiple \preprint commands are allowed.
%\preprint{}

\title{Influence of plasma particle flow on dust grain charging and on particle number density} %Title of paper

% repeat the \author .. \affiliation  etc. as needed
% \email, \thanks, \homepage, \altaffiliation all apply to the current author.
% Explanatory text should go in the []'s, 
% actual e-mail address or url should go in the {}'s for \email and \homepage.
% Please use the appropriate macro for the type of information

% \affiliation command applies to all authors since the last \affiliation command. 
% The \affiliation command should follow the other information.

\author{L. B. De Toni}
\email[]{luan.toni@ufrgs.br}
%\homepage[]{Your web page}
%\thanks{}
%\altaffiliation{}
\affiliation{Instituto de F{\'i}sica, Universidade Federal do Rio Grande do Sul, CP 15051, 91501-970, Porto Alegre, RS, Brazil}

\author{L. F. Ziebell}
\email[]{luiz.ziebell@ufrgs.br}
\affiliation{Instituto de F{\'i}sica, Universidade Federal do Rio Grande do Sul, CP 15051, 91501-970, Porto Alegre, RS, Brazil}

\author{R. Gaelzer}
\email[]{rudi.gaelzer@ufrgs.br}
\affiliation{Instituto de F{\'i}sica, Universidade Federal do Rio Grande do Sul, CP 15051, 91501-970, Porto Alegre, RS, Brazil}

% Collaboration name, if desired (requires use of superscriptaddress option in \documentclass). 
% \noaffiliation is required (may also be used with the \author command).
%\collaboration{}
%\noaffiliation

\date{\today}

\begin{abstract}
This study explores the dynamic evolution of dust electrical potential and plasma particle number densities with a focus on the charging of dust grains through electron and ion absorption, as described by the orbital motion limited (OML) theory. The initial model, which does not account for plasma particle sources and sinks, predicts that dust grains could eventually absorb all plasma particles, leading to a null electrical potential. To address this, we introduced source and sink terms considering  a finite region of space in order to simulate real conditions. Our findings indicate that, with the inclusion of plasma particle flow into and out of the region, dust grains reach a stable, non-zero equilibrium potential and the electron and ion densities reach an equilibrium value. This equilibrium is dependent on the size of the region; larger regions result in lower plasma densities and more negative equilibrium potentials. For extensive regions, the dust potential initially mirrors the scenario without sources or sinks but eventually deviates, showing increasing negative values as the region size grows. This behavior is attributed to the electron source term surpassing the combined sink and absorption terms at certain intervals along time evolution.
\end{abstract}

% insert suggested PACS numbers in braces on next line

\maketitle %\maketitle must follow title, authors, abstract and \pacs

% Body of paper goes here. Use proper sectioning commands. 
% References should be done using the \cite, \ref, and \label commands
\section{Introduction}

Dusty plasmas are common in astrophysical and laboratory environments. They present several properties which are different from those of conventional plasma since the dust grains acquire an electrical charge and change the collective behavior of the system \citep[][]{shukla_book}. Although there are many processes by which dust particles acquire electrical charge, one of the most important in astrophysical environments is the absorption of plasma particles. Inelastic collisions between plasma particles and dust grains result in the absorption of the particles' electrical charge by the dust. As a result, the dust particles become negatively charged simply because their encounters with the swift electrons are more frequent than with the ions \citep[][]{Spitzer1941}.

In addition to the absorption of plasma particles, dust grains may acquire an electrical charge due to photoelectric charging by electromagnetic radiation, secondary electron emission, field emission, among others. The charge and potential of the dust particle immersed in a plasma will be ultimately determined by the balance between the electron and ion current towards the dust particle’s surface \citep[][]{Khrapak2009,Shukla2009}.

The source of dust particles in space environments can be diverse. In planetary environments, such as the Earth's magnetosphere, interplanetary dust particles from the solar wind may be captured by the planet's gravity \citep[][]{Mendis1984}. Also, the dust flux occasionally increases in passages of meteoroids and comets closer to the Sun \citep[][]{STAUBACH1997,sykes2004interplanetary}.

Inside the heliosphere, dust particles have been detected through infrared emissions, particularly during solar eclipses \citep[][]{Lena1974,mankin1974coronal}, as well as by direct observations from missions like NASA’s Parker Solar Probe \citep[][]{howard2019near}. 

The existence of dust grains in stellar winds of distant stars has also been observed. Circumstellar dust in R Coronae Borealis (RCB) stars causes their brightness to drop up to 8 mag in visible light due to the presence of dust in distances as close as 1.5 stellar radii \citep[][]{Clayton1992}. In carbon-rich stars, carbon dust particles play an important role in the acceleration mechanism of their stellar winds, causing the lost of a notable amount of their masses by way of these powerful particle flows \citep[][]{Knapp1987_MassLoss,Mattson2011,Lieb_2025}.

When studying the characteristics of these dusty plasmas, it is common to consider the absorption of plasma particles as the main dust charging process \citep[][]{dejuli_schneider_1998,dejuli_2005,Ziebell_2005}. Although it is well known that other mechanisms may also play an important role in the plasma features, such as the effect of photoionization in the propagation and damping of electromagnetic waves \citep[][]{detoni2021}.

\citet{Allen_1957} developed a charging theory based on radial motion (ABR theory). It assumes that the plasma ions are collisionless and cold, so that ions initially far from the grain simply fall in straight radial lines onto the grain through its attractive electric field. ABR theory models the charging of a sphere in a plasma with cold ions, leaving aside the question of how a sphere charges in a plasma with ions of non-negligible kinetic energy.

To address the issues with ABR theory, one may use the most popular charging model, the orbit motion limited (OML) theory of \citet{MottSmith1926}, which takes into account the kinetic motion of the ions. The OML theory is known to predict accurately the dust potential for a variety of applications, despite its simplifying assumption of collisionless ion orbit that misses the effect of absorption radius around the dust surface \citep[][]{Allen_1992}.

More sophisticated models, such as the orbit motion (OM) theory, can account for absorption radii and generate expressions for dust potential and ion density as a function of distance from the grain \citep[][]{Bernstein1959}. Naturally, this comes with a more involved and less tractable algebra and calculus than those of the OML model, and it does not produce closed-form expressions for the dust potential.

However, the simplicity of the OML theory comes with inherent limitations. One of its key assumptions is that a plasma particle will be absorbed if its trajectory, dictated by the Coulomb potential, intersects the dust surface. Some studies \citep[][]{Weingartner_2001, GODENKO20235142} propose a model in which an electron reaching the grain surface has a certain probability of being reflected or transmitted with enough energy to escape. Additionally, an incoming electron may recombine with an ion on the grain surface. Moreover, the electrical potential equation can be modified to incorporate the ``image potential'', which arises from the polarization of the grain induced by the Coulomb field of the charge \citep[][]{Draine+1987}. Nevertheless, the OML model has proven to fit well with observations in many cases \citep[][]{Kimura_1998,Misra+2012,Yaroshenko2014}. Thus, we adopt it in this work to study the effects of the dusty cloud region on grain potential and plasma densities.

Several works provide a comprehensive analysis of ionization-recombination processes and dust charging dynamics and highlight the interplay between plasma absorption and other charging processes as key mechanisms determining the charge and sign of grains both in laboratory and space plasmas \cite{Fortov+2005,Draine2011}. When the dust density is high, a considerable amount of the available electrons and ions is bound to the dust. Thus, the density of free plasma particles in the plasma might be drastically reduced due to the presence of the dust, which in turn influences the charging behavior of the dust \citep[][]{Havnes+1987}.

One example of plasma particle depletion by dust grains can be found in noctilucent clouds. It is well understood that these regions consist of water ice particles with typical radii of $10$-$100$ nm and concentrations of $10$-$5000\,\text{cm}^{-3}\,$ \citep[][]{Hervig+2001,Baumgarten+2008}. These ice particles cause significant depletion of the electron density by electron attachment. One of the very first examples of such disturbances was the measurement of \citet{Pedersen+1970}; these authors reported a pronounced decrease in the electron density profile at an altitude of $87$ km. Moreover, ion depletion was observed in noctilucent clouds, which can be explained by the presence of at least \( 10^5 \,\text{cm}^{-3} \) ice particles \citep[][]{Balsiger+1996}.

Moreover, data from multiple instruments aboard Cassini revealed distinct plasma and dust regions around Enceladus, where dust particles interact with the surrounding plasma to establish equilibrium. These particles predominantly acquire negative charges \cite{ENGELHARDT2015453}, though a small fraction may become positively charged \cite{Yaroshenko+2014}. The likelihood of this effect primarily depends on the dust-to-plasma number density ratio. Notably, in both the E ring and the plume, electron densities are generally much lower than ion densities \cite{Morooka+2011,Farrell+2009}.

Since most space plasmas are in a dynamical environment, it may be considered that a flow of plasma particles in the system maintains the plasma species densities practically constant. In view of that, several studies on the potential of dust grains in a plasma do not take into account the dynamical variation on the plasma species number densities in the time period when the dust grains are being charged, calculating the equilibrium value considering constant electron and ion densities that satisfy the quasi-neutrality condition \citep[][]{Horanyi1996,Kimura_1998,Yaroshenko2014}.

However, if one wishes, e.g., to study the dynamical variation of the plasma distribution functions as the dust grains are charging, it is necessary to consider the changes of densities in each interval of momentum and position space. Therefore, it is necessary to write equations that describe the evolution of the distribution functions of plasma particles.

In the work of \citet{Basha_1989}, the charging time for grains is studied theoretically with some selected parameters to match the environment of space plasmas. The analysis shows the charging time is a strong function of plasma electron temperature and that the increasing density of the grains has an effect on the grain potential difference due to the depletion of the electron population in the plasma.

The work of \citet{Ostrikov+2001} studies the process of dust-charge variation considering the background density fluctuations caused by electron capture and release from dust grains. The analysis reveals that when dust charge and density are sufficiently high, variations in background electron density play a significant role in dust-charge relaxation.

In this work, we begin to explore the changes in the distribution functions related to the dust absorption of plasma particles, considering only the absorption of particles as the dust charging mechanism, using the OML theory. As a first approach, we study the evolution of the average densities together with the potential of a dust population, disregarding diffusion effects. Naturally, in a physical system, spatial variations in number density will inevitably arise as a consequence of the processes considered, leading to different equilibrium electric potentials for dust particles at various positions within the dust cloud. However, in this study, our focus is solely on analyzing the temporal variation in plasma densities, so we neglect diffusion terms in the model.

We see that considering only the absorption of particles described by the OML theory, the plasma densities and the dust electrical potential tend to zero after a large amount of time. This occurs because, as we will show, within the OML theory, a single grain could indefinitely absorb ions and electrons over time. Addressing this would require modifying the OML model to impose a limit on the number of particles a dust grain can absorb or allowing the absorbed plasma particles to be easily re-emitted back into the plasma. However, our goal is to study the time evolution of the dust potential and plasma densities using the OML model, given its widespread use in the dusty plasma literature.

To address this issue, we propose a simple model for the flow of plasma particles into and out of a finite region. As we will see, the equilibrium dust potential and plasma densities will ultimately depend on the size of the dusty plasma region.

This paper is written as follows: in section \ref{sec:charging} we introduce the dust charging model and present the potential and densities evolution equations; in section \ref{sec:sourcemodel} we derive the source and sink terms for the plasma particles; section \ref{sec:results} shows the numerical results for the model; finally, the conclusions are presented in section \ref{sec:conclusions}.

\section{The charging of dust grains}\label{sec:charging}
For the model of dust charging, we consider a homogeneous dusty plasma composed of electrons, ions, and spherical dust grains with constant radius $a$ and variable electrical charge $q_\mathrm{d}$. The assumption of a homogeneous plasma is commonly used to simplify both analytical and numerical analyses of plasma behavior. In a dusty plasma, inhomogeneities can lead to variations in equilibrium dust charges depending on the grain's position within the plasma. Therefore, this simplification does not account for these variations and assumes that all grain charges behave identically.

The charging occurs due to the absorption of plasma particles via inelastic collisions, and is described as an electric current incident on the grain's surfaces as follows \citep[][]{dejuli_schneider_1998}:
\begin{equation}
    I_\beta(q_\mathrm{d})=\frac{q_\beta}{m_\beta}\int d^3p \sigma_\beta(q_\mathrm{d},p) p f_\beta,
    \label{eq:I_beta}
\end{equation}
where $q_\beta$, $m_\beta$ and $f_\beta$ are, respectively, the charge, mass and distribution function of the plasma species $\beta$, $p$ is the momentum of the plasma particles, and
\begin{equation}
    \sigma_\beta(q_\mathrm{d},p)=\pi a^2 \left(1-\frac{2q_\mathrm{d}q_\beta m_\beta}{ap^2}\right) H\left(1-\frac{2q_\mathrm{d}q_\beta m_\beta}{ap^2}\right)
\end{equation}
is the collisional cross section, with $H(x)$ denoting the Heaviside function. The quantity $I_\beta(q_\mathrm{d})$ corresponds to the fraction of the total current on the grain's surface due to the $\beta$ species.

This charging model is derived from the OML theory and does not take into account the presence of an external magnetic field. Such a field would change the path described by electrons and ions (in this case, cyclotron motion around the magnetic field lines). However, for weakly magnetized dusty plasmas, it has been shown \citep[][]{Chang_1993,Kodanova+2019} that this model is valid when the relation $a \ll r_{Le}$ is satisfied, with $r_{Le}$ being the electron Larmor radius.

This theory assumes that electrons and ions are absorbed when their trajectories intersect the surface of the grain, meaning that the limiting impact parameter corresponds to a trajectory tangential to the grain’s surface. These trajectories are collisionless, implying that other plasma particles do not affect their motion near the dust grain, not even other charged dust grains. Thus, this approximation is valid when \( a \ll \lambda_D < \lambda_{mfp} \), where $\lambda_D$ is the Debye length and \( \lambda_{mfp} \) is the mean free path for ion and electron collisions with dust particles. This approximation is useful because it allows the absorption cross-section of the particles to be calculated using the conservation laws of energy and angular momentum, regardless of the complexity of the electric potential near the dust grain.

Considering only the absorption of plasma particles by the grains, the time evolution of the dust charge is given by
\begin{equation}
    \frac{dq_\mathrm{d}}{dt}=\sum_\beta I_\beta(q_\mathrm{d}).
    \label{eq:dqdt}
\end{equation}

We now write equations \eqref{eq:I_beta} and \eqref{eq:dqdt} in terms of the following dimensionless variables:
\begin{equation}
\begin{alignedat}{2}
    &\tau=\frac{t}{\tau_c},\quad \psi_{d}=\frac{eq_\mathrm{d}}{ak_{B}T_{e}},\quad  \chi_{\beta\mathrm{d}}=\frac{q_{\beta}q_\mathrm{d}}{ak_{B}T_{\beta}},\quad \chi_{e}=\frac{e^{2}}{ak_{B}T_{e}}, \\
    & \varepsilon_\beta=\frac{n_{\beta}}{n_0}, \quad\varepsilon_\mathrm{d}=\frac{n_\mathrm{d}}{n_0}, \quad \tilde{I}_\beta = \frac{I_\beta}{2\sqrt{2\pi}a^2en_0 v_{Te}},
    \label{eq:dimensionless_var}
\end{alignedat}
\end{equation}
where
\begin{equation}
    \tau_c=\sqrt{2\pi}\frac{\lambda_{De}^2}{av_{Te}}
    %\tau_{c}=\frac{1}{2\sqrt{2\pi}}\frac{(m_e k_{B}T_{e})^{1/2}}{a n_{0}e^{2}}
\end{equation}
is the characteristic charging time, $k_B$ is the Boltzmann constant, $T_\beta$, $n_\beta$ and $v_{T\beta}$ are the temperature, density and thermal velocity of the plasma species $\beta$, $e$ is the elementary charge, $n_0$ is the initial number density of plasma particles and
\begin{equation}
    \lambda_{De}=\left( \frac{k_B T_e}{4\pi n_0 e^2} \right)^{1/2}
\end{equation}
is the initial electron Debye length.

In terms of these variables, and considering that the $f_\beta$ are described by Maxwellian distributions, the equation for the dust charge evolution is written as
\begin{equation}
    \frac{d\psi_\mathrm{d}}{d\tau}=\sum_{\beta}\tilde{I}_{\beta}
    \label{eq:dpsi_d}
\end{equation}
with
\begin{align}
    &\tilde{I}_{e}=-\epsilon_{e}
    \begin{cases}
        \exp\left(-\chi_{e{\mathrm{d}}}\right) & \psi_\mathrm{d}<0\\
        \left(1-\chi_{e\mathrm{d}}\right) & \psi_\mathrm{d}\geq0
    \end{cases}, \\
    &\tilde{I}_{i}=\epsilon_{i}\frac{v_{Ti}}{v_{Te}}
    \begin{cases}
        \left(1-\chi_{i\mathrm{d}}\right) & \psi_\mathrm{d}\leq0\\
        \exp\left(-\chi_{i\mathrm{d}}\right) & \psi_\mathrm{d}>0
    \end{cases}.
\end{align}

The Maxwellian velocity distribution is the most commonly used probability distribution, as it assumes that a system of charged particles has reached thermodynamic equilibrium. However, we point out that in many space environments, plasmas may deviate from the Maxwellian distribution due to their often collisionless nature. In such regions, the plasma system may exhibit superthermal and nonthermal characteristics. These deviations have been observed in the magnetospheres of Jupiter and Saturn \cite{Mihalov_2000,porco2005cassini}, Earth's magnetosphere \cite{Runov+2015}, the vicinity of the Moon \cite{Futaana+2003}, and the solar wind \cite{Wilson_2019,Bercic_2019,Bercic_2020}. Despite these deviations, we choose to work with the Maxwellian distribution as it provides simple analytical solutions and our focus will be on studying the time variations of the averaged distributions in momenta.

To numerically solve these equations, we consider the following plasma parameters, which are used throughout this paper: $n_{e0}=n_{i0}=n_0=10^9\,\text{cm}^{-3}$, $T_e=T_i=10^4\,\text{K}$, $a=10^{-4}\,\text{cm}$ and $\varepsilon_\mathrm{d}=10^{-5}$. These parameters are typical of stellar winds coming from carbon-rich stars \citep[][]{tsytovich_2004}.

\begin{figure}
    \centering
    \includegraphics[width=\columnwidth]{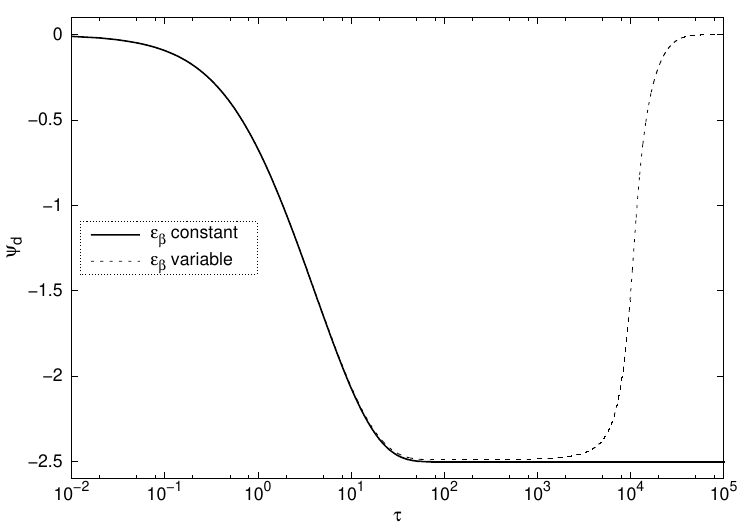}
    \caption{Time evolution of the dimensionless dust potential considering constant plasma particles densities (continuous lines) and variable densities (dashed lines), following equations \eqref{eq:deps_i} and \eqref{eq:deps_e}. Plasma parameters are as follows: $n_{e0}=n_{i0}=n_0=10^9\,\text{cm}^{-3}$, $T_e=T_i=10^4\,\text{K}$, $a=10^{-4}\,\text{cm}$ and $\varepsilon_\mathrm{d}=10^{-5}$.}
    \label{fig:evol_nosource}
\end{figure}

When we consider that the plasma densities are unaffected by the dust absorption (i.e., $\varepsilon_e=\varepsilon_i=1$), due to some source of particles that compensate the ones absorbed by the grains, we get the result depicted by the continuous line in Fig. \ref{fig:evol_nosource}. The grain will show negative values of the normalized electrical potential $\psi_\mathrm{d}$, given that the collisions with electrons are more common. After a certain value of the time $\tau$, the grains achieve an equilibrium potential value; at this point, the sum of the currents on the grains' surface is equal to zero.

Alternatively, we may consider that the plasma densities are affected as the particles are absorbed by the dust particles, without any source adding electrons and ions to the system. From the quasi-neutrality condition,
\begin{equation}
    \sum_\beta n_\beta q_\beta + n_\mathrm{d}q_\mathrm{d}=0,
\end{equation}
we see that the variation in the dust charge will affect the plasma densities. Taking the derivative with respect to time, using equation \eqref{eq:dqdt} and the dimensionless variables \eqref{eq:dimensionless_var}, we get
\begin{equation}
    \left( \frac{d\varepsilon_i}{d\tau} + \varepsilon_\mathrm{d} \frac{1}{\chi_e} \tilde{I}_i \right) + \left( -\frac{d\varepsilon_e}{d\tau} + \varepsilon_\mathrm{d} \frac{1}{\chi_e} \tilde{I}_e \right) =0.
    \label{eq:quasineut_deriv}
\end{equation}
Since the ion and electron currents on the grains' surface will affect only the corresponding species density, we consider that the sums within each parenthesis in equation \eqref{eq:quasineut_deriv} are equal to zero, so we write for the densities evolution:
\begin{align}
    &\frac{d\varepsilon_i}{d\tau} =- \varepsilon_\mathrm{d} \frac{1}{\chi_e} \tilde{I}_i,\label{eq:deps_i}\\
    &\frac{d\varepsilon_e}{d\tau} = \varepsilon_\mathrm{d} \frac{1}{\chi_e} \tilde{I}_e.\label{eq:deps_e}
\end{align}
In this scenario, plasma neutrality is preserved because any change in species density is balanced by a corresponding change in the dust electrical charge, without the creation or destruction of additional particles.

\begin{figure}
    \centering
    \includegraphics[width=\columnwidth]{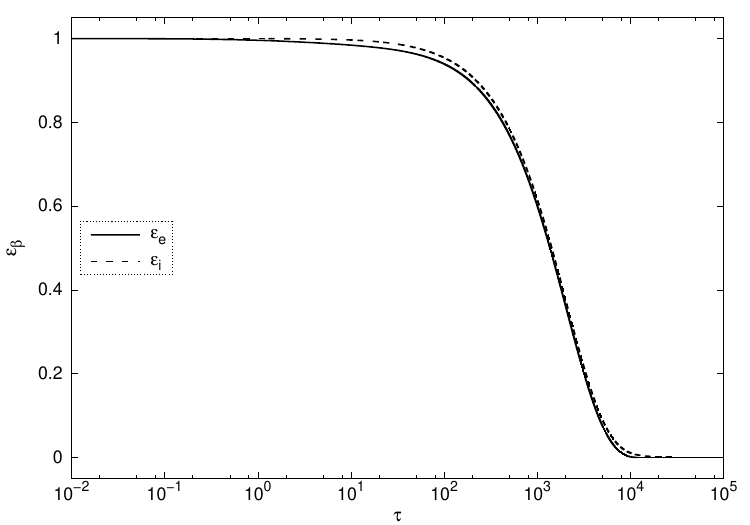}
    \caption{Time evolution of the densities $\varepsilon_\beta=n_\beta/n_0$ for electrons and ions following equations \eqref{eq:dpsi_d}, \eqref{eq:deps_i} and \eqref{eq:deps_e} and same plasma parameters as in Fig. \ref{fig:evol_nosource}.}
    \label{fig:dens_nosource}
\end{figure}

The solution of the equations \eqref{eq:dpsi_d}, \eqref{eq:deps_i} and \eqref{eq:deps_e} is depicted by the dashed line in Fig. \ref{fig:evol_nosource} for the electric potential evolution, and in Fig. \ref{fig:dens_nosource} for the densities evolution, using the same parameters as before. We notice that the electron and ion densities will diminish constantly with respect to time. Since there is no source of plasma particles and the absorption model \eqref{eq:I_beta} does not have any restriction to how many particles the dust may absorb, the grains will absorb the particles continuously such that, for a sufficiently large time, there will be no more particles in the plasma.

As a consequence, the electrical potential in Fig. \ref{fig:evol_nosource} will firstly diminish because of the more frequent electron collisions. However, at a given time, the ion current becomes larger and the potential starts to increase towards zero. After a large amount of time, the grains will have absorbed all the plasma particles and their potential will go back to zero, which is the equilibrium value in this case.

However, we note that when the plasma densities decrease to the point where the condition \( \lambda_D < \lambda_{mfp} \) is no longer satisfied, the OML theory becomes invalid. In this case, the assumption that plasma particle trajectories remain unaffected by other dust grains in the system no longer holds. Under these conditions, the absorption model should be refined to account for the electric potential of multiple dust particles, and also to consider that it deviates from a purely Coulombian form at significant distances from the dust surface. Nevertheless, we believe it is valuable to investigate the conditions under which the OML theory remains valid or breaks down and to analyze the system’s response in such scenarios.

In space plasmas, the environment is not static and generally there is a flow of plasma particles originated, e.g., from the stellar wind \citep[][]{Kislyakiva2014}. Additionally, the OML model for plasma particle absorption does not limit the amount of particles that can be absorbed by the dust grains, and it will continue to happen as long as there are particles present in the system. Therefore, the issue presented when we study the density evolution together with the dust potential must be addressed by considering a flow of particles into the plasma or/and improving the absorption model used. Given the widespread use in the literature of the OML theory, we use the former approach as explained in the next section.

\section{Model for source and sink of particles in a finite region}\label{sec:sourcemodel}

\begin{figure}
    \centering
    \includegraphics[width=.7\columnwidth]{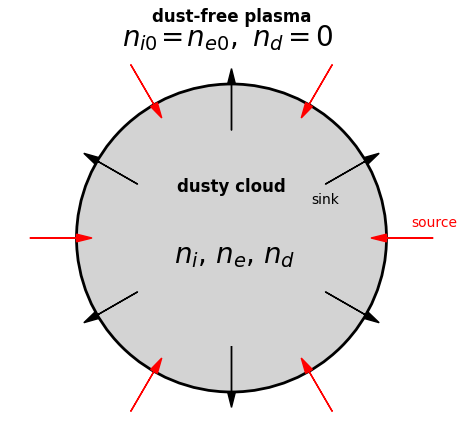}
    \caption{Schematic diagram of the modeled plasma.}
    \label{fig:diagram}
\end{figure}

The approach we use in this work considers a finite region where dust grains are present and, consequently, alter the plasma densities within it. This region is surrounded by a dustless plasma with constant density, which is the same as the initial density of the region considered. The plasma particles at the borders of that region will flow in and out; this flow will be proportional to the density and thermal velocity of that species. A schematic diagram of the modeled plasma configuration is depicted in Fig. \ref{fig:diagram}. As a first approach, we simplify the model considering that the particles will flow in and out along all volume of the region and not only at the borders, not considering the diffusion term that appears in the fluid equations.

As a result, this region will become electrically charged, as expected for a finite-sized dusty region embedded in an infinite plasma \citep[][]{Havnes+1987}. The approach utilized takes into account the dynamical evolution of the plasma particles number densities and the associated diffusion of particles in and out of the finite region. However, it does not take into account the electric potential acquired by the finite region or the consequent acceleration or deceleration of plasma particles entering or leaving it.

Therefore, we consider that the density outside the dusty plasma region is $n_0$ and there is a flow of plasma species $\beta$ into that region of $n_0 v_{T\beta}$ particles per unit area per unit time. Multiplying this term by the region surface area $A_s$ will result in the total number of particles flowing into the region. We also divide this term by the region volume $V$ to result in the density flow of particles per unit time. Hence, we write the source of plasma particles as
\begin{equation}
    \frac{dn_\beta}{dt}\bigg|_{\text{source}}=n_0 v_{T\beta} \frac{A_s}{V}.
\end{equation}

Similarly, we write a model for the flow to the outside of the region, which will be proportional to the density $n_{\beta}(t)$. This flow is variable in time since the density inside the region will change due to the capture of plasma particles by the dust grains. We call this term the sink of plasma particles. Although the absorption of plasma particles by the dust grains also configures a sink of particles, we refer to this term as the `absorption' term, and the flow of particles to the outside of the region as the `sink' term. Therefore, we write the sink term as
\begin{equation}
    \frac{dn_\beta}{dt}\bigg|_{\text{sink}}=-n_\beta(t) v_{T\beta} \frac{A_s}{V}.
\end{equation}

Considering a spherical region of radius $R$ and the inter-grain distance as $r_d=1/n_\mathrm{d}^{1/3}=1/(\varepsilon_\mathrm{d} n_0)^{1/3}$, we write the source and sink terms in terms of the dimensionless variables \eqref{eq:dimensionless_var} as
\begin{align}
    &\frac{d\varepsilon_\beta}{d\tau}\bigg|_{\text{source}}=3\sqrt{2\pi}\frac{v_{T\beta}}{v_{Te}}\frac{\lambda_{De}^2}{a(R/r_d)}(\varepsilon_\mathrm{d} n_0)^{1/3}, \\
    &\frac{d\varepsilon_\beta}{d\tau}\bigg|_{\text{sink}}=-3\sqrt{2\pi}\frac{v_{T\beta}}{v_{Te}}\frac{\lambda_{De}^2}{a(R/r_d)}(\varepsilon_\mathrm{d} n_0)^{1/3}\varepsilon_\beta.
\end{align}
The absorption term is derived from equations \eqref{eq:deps_i} and \eqref{eq:deps_e}, which are based on the assumption that these equations were developed under the quasi-neutrality condition, without any sources or sinks of particles. Under this framework, changes in species densities are solely due to the absorption of plasma particles by the grains, i.e.,
\begin{equation}
    \frac{d\varepsilon_\beta}{d\tau}\bigg|_{\text{abs}}=-\text{sgn}(q_\beta)\varepsilon_\mathrm{d}\frac{1}{\chi_e}\tilde{I}_\beta.
\end{equation}
Therefore, considering all the mechanisms of the system, the total density variation is given by
\begin{equation}
    \frac{d\varepsilon_\beta}{d\tau}=\frac{d\varepsilon_\beta}{d\tau}\bigg|_{\text{abs}}+\frac{d\varepsilon_\beta}{d\tau}\bigg|_{\text{source}}+\frac{d\varepsilon_\beta}{d\tau}\bigg|_{\text{sink}}.
    \label{eq:depsdtau_total}
\end{equation}

\section{Numerical Analysis}\label{sec:results}

\begin{figure}
    \centering
    \includegraphics[width=\columnwidth]{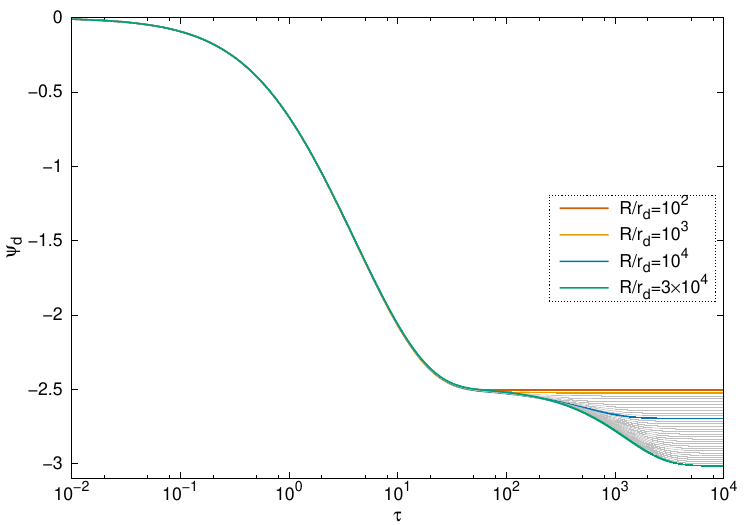}
    \caption{Time evolution of normalized dust electrical potential for several dust region radii $R$. Gray lines represent regions with normalized radius $R/r_\mathrm{d}$ with an increment of $1{\times}10^3$ in relation to the previous line. Same plasma parameters as in Fig. \ref{fig:evol_nosource}.}
    \label{fig:evol}
\end{figure}

Now we solve numerically equations \eqref{eq:dpsi_d} and \eqref{eq:depsdtau_total} for the same plasma parameters considered in Fig. \ref{fig:evol_nosource}. The time evolution of the normalized dust electrical potential for a given region size is depicted in Fig. \ref{fig:evol}. The colored lines show the curves corresponding to the ratio of the dusty region radius over inter-grain distance with values $R/r_\mathrm{d}=10^2,10^3,10^4$ and $3{\times}10^4$; while gray lines represent regions with normalized radius with an increment of $1{\times}10^3$ in relation to the previous line. Given the numerical parameters, these $R/r_d$ values correspond to spherical regions with radii ranging from a few centimeters to a few meters. 

We observe that when a finite size for the dusty plasma region is considered, with a source of plasma particles, the dust potential tends to achieve an equilibrium value $\psi_{\mathrm{d},\text{eq}}$. As the value of the region radii $R$ increases, the equilibrium potential tends to be higher in modulus. Furthermore, the time needed for the potential to achieve an equilibrium value also depends on the region size. As shown in table \ref{tab:eq_values}, the dust grains take longer to reach such value of $\psi_{\mathrm{d},\text{eq}}$ for higher values of $R/r_\mathrm{d}$.

\begin{table}
	\centering
	\caption{Numerical values for equilibrium dust potential $\Psi_{\mathrm{d},\text{eq}}$, electron density $\varepsilon_{e,\text{eq}}$ and ion density $\varepsilon_{i,\text{eq}}$ achieved in a time $\tau_\text{eq}$, for different dusty region radius $R/r_\mathrm{d}$.}
	\label{tab:eq_values}
	\begin{tabular}{cccccc} 
		  $R/r_\mathrm{d}$ & $\Psi_{\mathrm{d},\text{eq}}$ & $\tau_\text{eq}$ & $\varepsilon_{e,\text{eq}}$ & $\varepsilon_{i,\text{eq}}$ & \\
		\hline
		  $10^2$ & $-2.50591$ & $300$   & $0.99994$ & $0.99729$&\\
		$10^3$ & $-2.52436$ & $1413$  & $0.99938$ & $0.97340$&\\
		$10^4$ & $-2.69708$ & $12699$ & $0.99480$ & $0.77714$&\\
		$3{\times}10^4$ & $-3.01591$ & $27098$ & $0.98873$ & $0.51693$&\\
		\hline
	\end{tabular}
\end{table}

\begin{figure}
    \centering
    \begin{minipage}[c]{\columnwidth}
        \centering
        \includegraphics[width=\columnwidth]{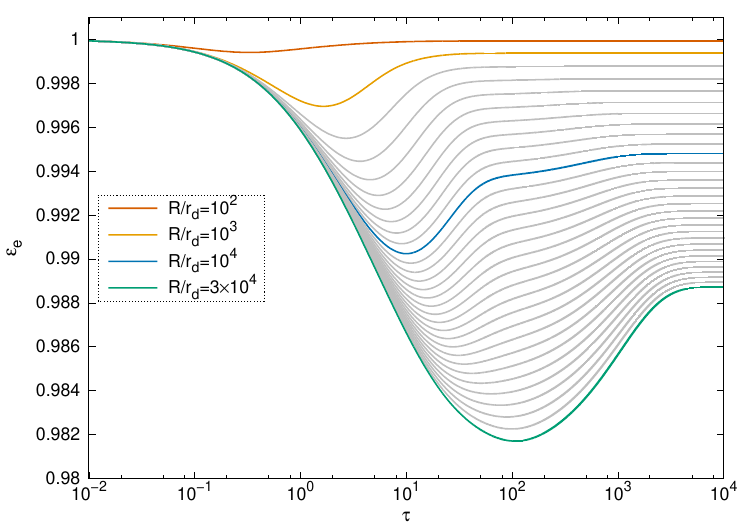}%
    \end{minipage}\vspace{\floatsep}
    \begin{minipage}[c]{\columnwidth}
        \centering
        \includegraphics[width=\columnwidth]{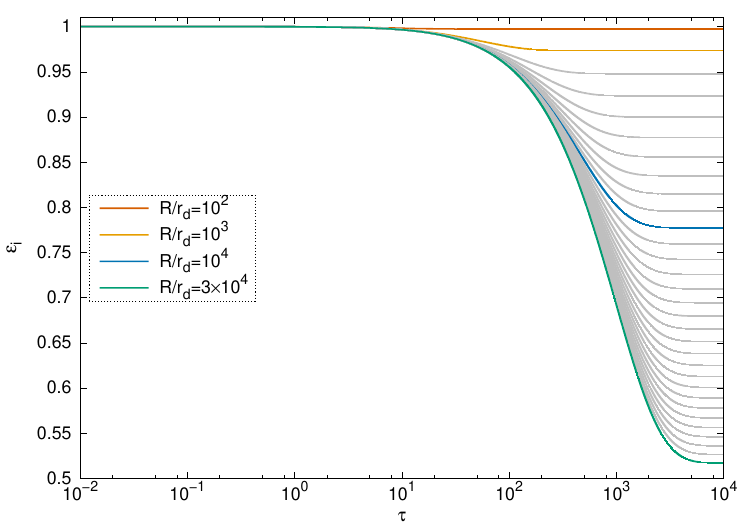}
    \end{minipage}
    \caption{Time evolution of electron (top panel) and ion (bottom panel) dimensionless densities $\varepsilon_\beta=n_\beta/n_0$ for several dusty region radii $R$. Gray lines represent regions with dimensionless radius $R/r_d$ increased by $1{\times}10^3$ in relation to the previous line.}
    \label{fig:evol_densities}
\end{figure}

In Fig. \ref{fig:evol_densities} we can see the time evolution of electron (top panel) and ion (bottom panel) densities. Both plasma species tend to achieve a lower equilibrium value for bigger regions. For the region sizes considered, the electron density decreases by no more than $2\%$ of its initial value. On the other hand, the ion number density is much more affected, with the value of equilibrium attaining nearly $50\%$ of the initial value, for $R/r_\mathrm{d}=3{\times}10^4$.

The ion densities show always decreasing values for higher time steps until the equilibrium value is numerically achieved. Meanwhile, the electron densities start decreasing in time and reach a minimum value; after this point, the densities increase to reach the equilibrium value. While the curves corresponding to $R/r_\mathrm{d}=10^2,10^3$ and $3{\times}10^4$ show always increasing values after the minimum point, we see in the $R/r_\mathrm{d}=10^4$ curve a fluctuation of the densities between approximately $\tau=10^2$ and $10^3$ before reaching the equilibrium value. 

This behavior can be seen in some of the gray curves, which show the results for an increment of $1{\times}10^3\,R/r_\mathrm{d}$ in relation to the previous line. These lines show a smooth transition between the colored lines and indicate that this fluctuation only happens in a given interval of $R$ values. These fluctuations likely result from the interplay between ion and electron currents on the grain surfaces and the flow of particles into and out of the considered region, since the model does not account for external forces or diffusion effects. We observe that this phenomenon occurs within a specific time interval, without any clear temporal pattern, as the dust eventually reaches its equilibrium potential. Interestingly, this effect is only present within a certain range of region sizes \( R \); it is not observed for significantly larger or smaller regions, or, if present, the variations are too minor to detect.

\begin{figure}
    \centering
    \includegraphics[width=\columnwidth]{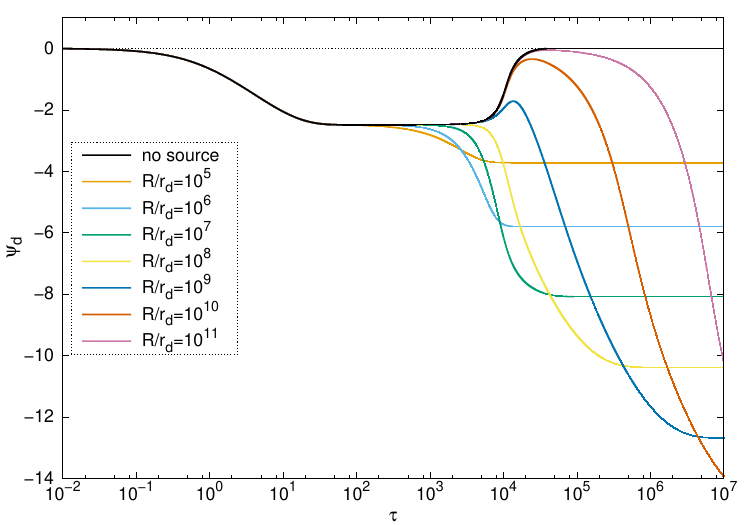}
    \caption{Time evolution of dimensionless dust electrical potential for several dusty region radii $R$. The black line represents the case without source or sink of plasma particles.}
    \label{fig:evolE5plus}
\end{figure}

\begin{figure}
    \centering
    \begin{minipage}[c]{\columnwidth}
        \centering
        \includegraphics[width=\columnwidth]{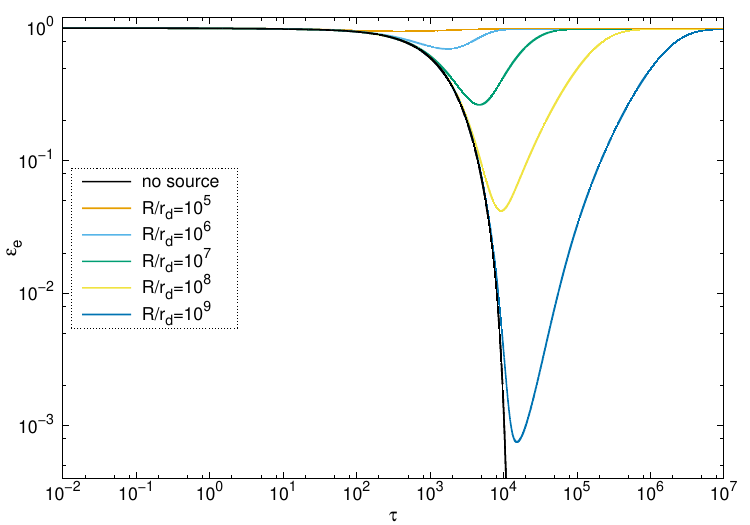}%
    \end{minipage}\vspace{\floatsep}
    \begin{minipage}[c]{\columnwidth}
        \centering
        \includegraphics[width=\columnwidth]{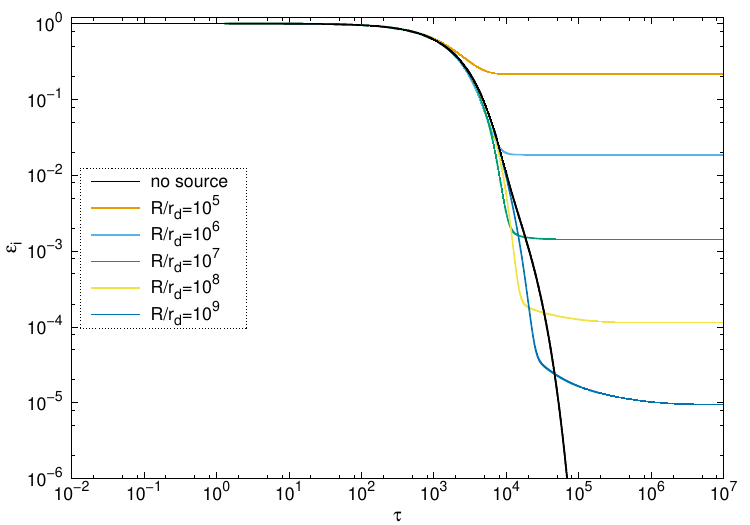}
    \end{minipage}
    \caption{Time evolution of electron (top panel) and ion (bottom panel) dimensionless densities $\varepsilon_\beta=n_\beta/n_0$ for several dusty region radii $R$. Black lines represent the cases without source or sink of plasma particles.}
    \label{fig:evolE5plus_densities}
\end{figure}

Now we look at larger regions, ranging from tens of meters to the order of $10^5$\,km, in Figs. \ref{fig:evolE5plus} and \ref{fig:evolE5plus_densities}. Fig. \ref{fig:evolE5plus} shows the time evolution of the dimensionless dust electrical potential for several $R/r_d$ values (colored lines) and for the case with no source or sink of plasma particles (black line), i.e., considering an infinitely large dusty region. As discussed before, the ``no source'' line will firstly show decreasing values of the dust electric potential, since the grains absorb more plasma electrons than ions. However, for higher time steps, the electron density decreases sufficiently for the ion absorption to become greater and the dust potential starts to increase in value. Eventually, the dust grains will absorb all the plasma particles and their electrical potential will go back to zero, since there is no source of electrons and ions.

We highlight that for these sizes of the dusty region, the plasma densities decrease to levels where the OML theory is no longer valid, as the Debye length exceeds the plasma-dust collisional mean free path. However, we find it valuable to illustrate how the model behaves under these extreme conditions and to examine the dynamics of its various terms.

Looking at the colored lines of Fig. \ref{fig:evolE5plus} we see that the lines tend to follow the `no source' case until a given time step. For higher values of $R/r_d$, the colored lines follow the black line during a longer time interval. If we tend the region radius to infinity, both the source and sink terms in equation \eqref{eq:depsdtau_total} would tend to zero and we would get back the case with no source or sink of plasma particles, as expected. However, we observe that after a given instant, the potentials stop approaching zero and start to become more negative, achieving an equilibrium value further away from zero as $R/r_d$ increases.

This effect is related to the source of plasma particles, as we can see in Fig. \ref{fig:evolE5plus_densities}. While the electrons and ions densities have ever decreasing values for the case without source or sink of particles (black lines), the colored lines show that when we consider a finite region, the electron density will start to increase after it reaches a minimum value, until it reaches its equilibrium. This rise in electron density, which can be of several orders of magnitude, causes the dust potential to reach high negative values.

\begin{figure*}
    \centering
    \begin{minipage}[t]{0.49\textwidth}%
        \centering
        \includegraphics[width=1\columnwidth]{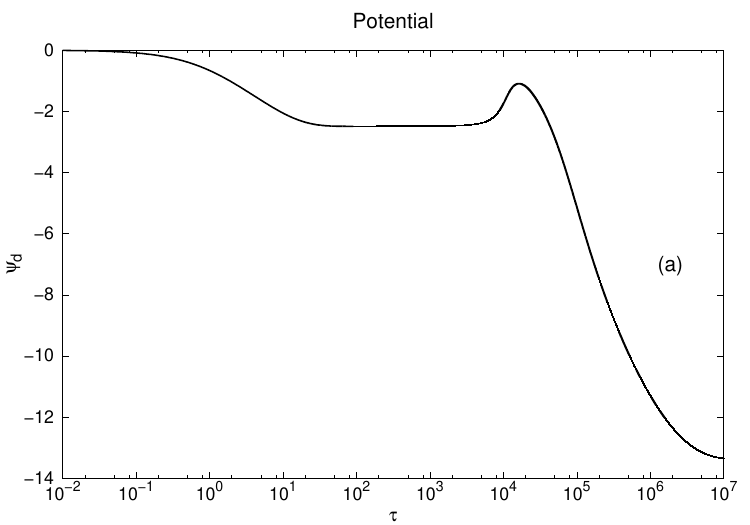}%
    \end{minipage}\hfill{}%
    \begin{minipage}[t]{0.49\textwidth}%
        \centering
        \includegraphics[width=1\columnwidth]{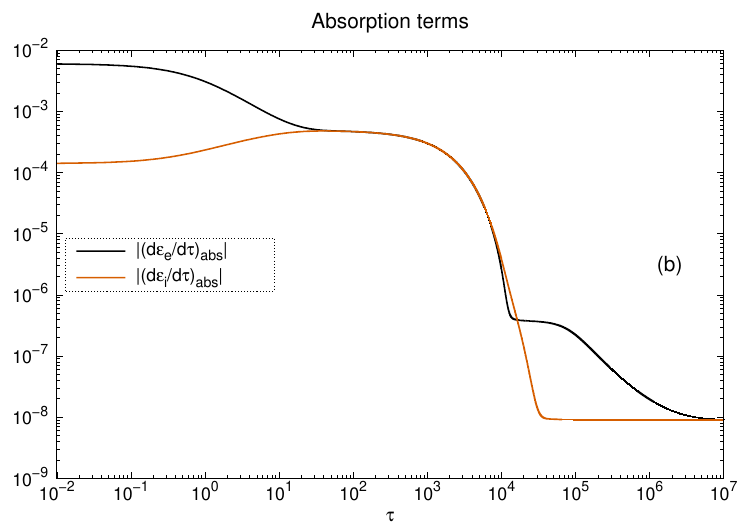}%
    \end{minipage}\vspace{\floatsep}
    \begin{minipage}[t]{0.49\textwidth}%
        \centering
        \includegraphics[width=1\columnwidth]{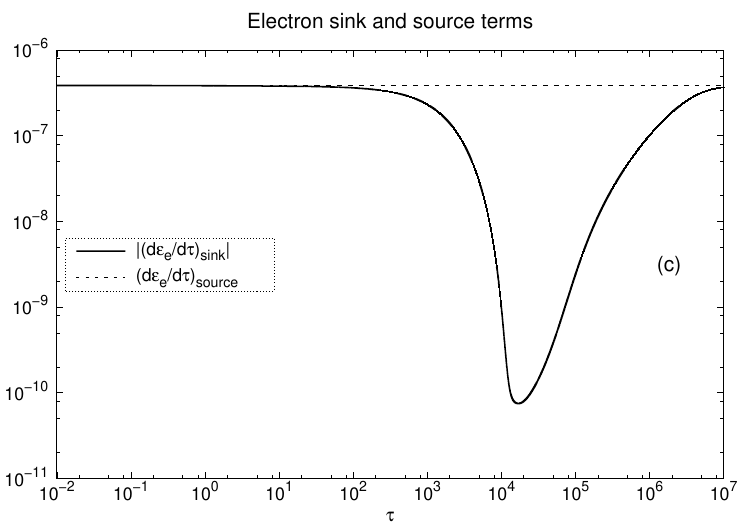}%
    \end{minipage}\hfill{}%
    \begin{minipage}[t]{0.49\textwidth}%
        \centering
        \includegraphics[width=1\columnwidth]{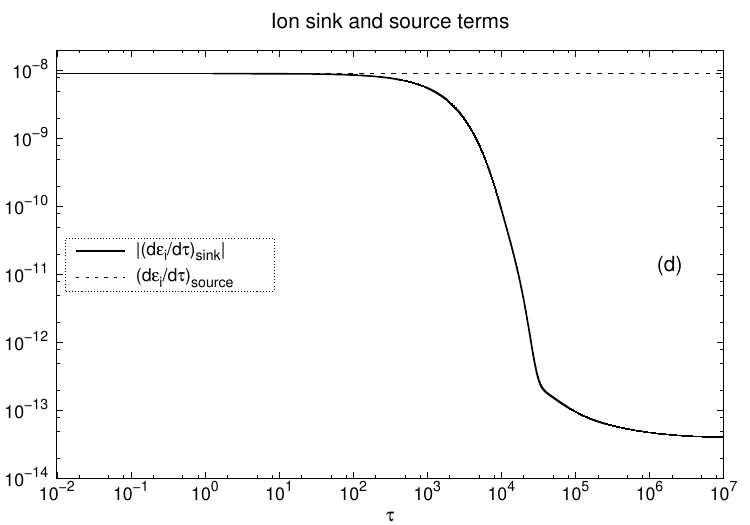}%
    \end{minipage}\vspace{\floatsep}
    \begin{minipage}[t]{0.49\textwidth}%
        \centering
        \includegraphics[width=1\columnwidth]{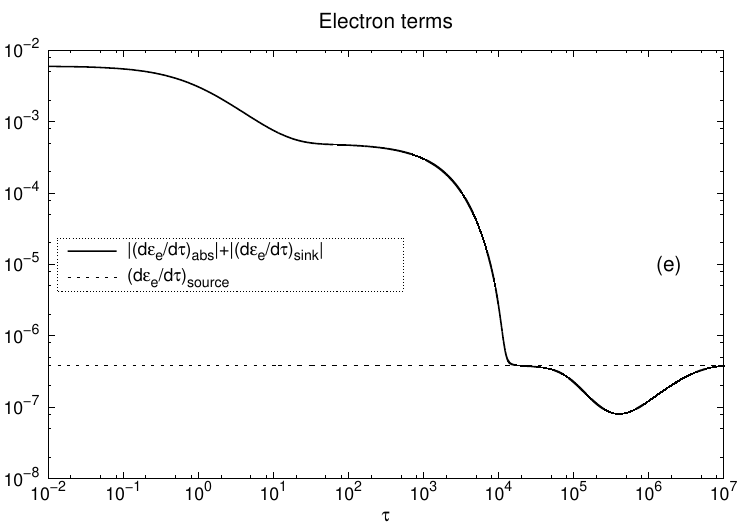}%
    \end{minipage}\hfill{}%
    \begin{minipage}[t]{0.49\textwidth}%
        \centering
        \includegraphics[width=1\columnwidth]{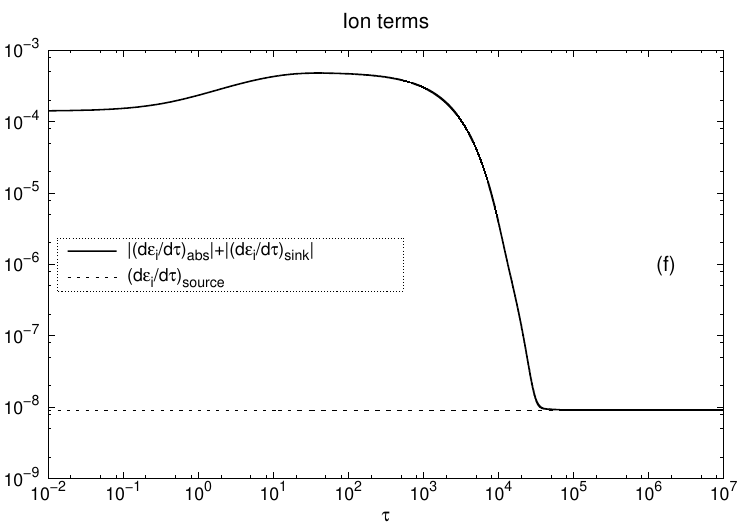}%
    \end{minipage}
    \caption{Time evolution of the absolute values of the terms in equation \eqref{eq:depsdtau_total}, for $R/r_d=10^9$. (a) Dust electrical potential. (b) Absorption terms for electrons and ions. Source (dashed lines) and sink (continuous lines) terms for (c) electrons and (d) ions. Source term (dashed lines) and total subtraction (continuous lines) of (e) electrons and (f) ions.}
    \label{fig:terms}
\end{figure*}

In order to better understand the dynamics involving the electric potential for large values of radii, we plot the absolute values of the different terms in equation \eqref{eq:depsdtau_total} separately for $R/r_d=10^9$ in Fig. \ref{fig:terms}. Panel (a) shows the evolution of the dimensionless electric potential. Panel (b) shows the absorption term $(d\varepsilon_\beta/d\tau)_\text{abs}$ in equation \eqref{eq:depsdtau_total} for electrons (black line) and ions (colored line). The source and sink terms, $(d\varepsilon_\beta/d\tau)_\text{source}$ and $(d\varepsilon_\beta/d\tau)_\text{sink}$, are shown in panel (c) for electrons and in panel (d) for ions. Panels (e) and (f) show again the source term together with the sum of the absorption and sink terms, i.e., the total subtraction of particles in the plasma, for electrons and ions, respectively.

As we look at the results for a large value of $R/r_d$, we observe that the absorption terms in panel (b) are a few orders of magnitude higher than the sink and source terms, in panels (c) and (d), for low $\tau$ values. At the beginning, the electron absorption term is higher than the ion term, making the dust potential to increase in negative values. As the time passes, both absorption terms are practically equal and the potential seems to achieve a stable value, however, the ion absorption term becomes higher and the potential starts to increase towards zero.

The electron and ion absorption terms then start decreasing together with their densities, however, around $\tau=10^4$, the electron source term becomes higher than the sum of the absorption and sink term, as seen in panel (e), i.e., the electron density starts increasing and the absorption term stops decreasing. Not long after that, we see that the electron absorption term is once again higher than the ion absorption term, and the electrical potential starts moving towards higher negative values. Since the potential becomes more negative, the electron collision becomes more difficult and this term starts decreasing again until it reaches the same value of the ion absorption term, at this point the dust potential is at equilibrium.

As we see, the analysis of the electric potential dynamics reveals significant insights into the behavior  of absorption, source and sink terms as modeled in equation \eqref{eq:depsdtau_total}. There is a strong interdependence between the dominance of each term and the instant value of the dust potential and the plasma densities.

\section{Conclusions}\label{sec:conclusions}

We studied the time evolution of the dust electrical potential and plasma particles densities considering that dust grains are charged by the absorption of electrons and ions. 

The initial model for the absorption of plasma particles is proposed following the OML theory. We noticed that this absorption model does not limit the amount of plasma particles that can be absorbed by the grains, so that, if we do not consider a source of electrons and ions, the dust grains will eventually absorb all the plasma particles of the system and their electrical potential will be null. 

To address this issue, we modeled simple expressions for the source and sink of plasma particles in a finite region. Taking into account the flow of plasma particles into and out of the considered dusty region, the dust grain potential and plasma particle densities are able to achieve non-zero equilibrium values. These equilibrium values will depend on the region size. As the region radii increase, the equilibrium plasma densities are lower, and the equilibrium potential tends to be more negative.

For larger regions, the time evolution of the dust potential tends to follow the case where no source nor sink of plasma particles is considered. However, after a certain time period, the potential curves deviate from the `no source' case and start showing increasing values in modulus, with the equilibrium values of the potential further from zero as the radius increases. To better understand this effect, we looked at the different terms in the time evolution equation for plasma density and saw that at a given time step the electron source term becomes higher than the sum of the sink and absorption terms, causing the electron density and absorption to increase, taking the potential to high negative values.

A comparison of our results with experimental findings would be valuable. The investigations by \citet{Goertz+2011} aim to provide a comprehensive study of dust charge variability in dense dust clouds embedded within a dust-free plasma, using a combination of probe techniques, including Langmuir and emissive probes. Their findings confirmed that electron and ion densities decrease inside the dust cloud, consistent with previous experiments \citep[][]{Trottenberg+2003}. Our numerical results reproduce this effect. Other experiments \citep[][]{Douglass+2011} indicate that the dust surface potential varies throughout the plasma due to spatial fluctuations in ion and electron flow, which result from density variations. This effect is not accounted for in our model.

Our objective in this work was to develop a model that incorporates the OML theory, given its widespread use in the literature. In cases where the density decreases substantially and the theory loses its validity, further research is needed to establish a more comprehensive framework. Additionally, the issue of continuous absorption of plasma particles by dust grains could be addressed within the absorption model. A more realistic approach would involve, e.g., the recombination of electrons and ions on the surface of a dust particle, followed by the re-ionization of the released neutrals to maintain an equilibrium plasma density. Furthermore, the model could consider a sticking coefficient, i.e., when an electron reaches the grain surface, it has some probability of being reflected from (or transmitted through) the grain with enough energy to escape \citep[][]{Weingartner_2001}.

One limitation of the model is that it does not account for diffusion effects. In reality, particles should enter the considered region through its borders rather than being uniformly distributed throughout the volume, as currently assumed. This would lead to spatial variations in plasma density, which, in turn, would influence the charge of grains at different positions within the region. Specifically, variations in plasma density around a particle would alter its charging currents. Additionally, the model does not consider the effects of the finite region itself being electrically charged, such as the acceleration of plasma particles entering or leaving the region. Addressing these limitations in future work could help develop a more comprehensive model.

Also, a model that incorporates the spatial distribution of charges along with electromagnetic field effects would enable the study of how wakefield structures influence overall plasma dynamics. These structures could accelerate electrons, impacting the charging currents and, consequently, the evolution of the dust potential.

The investigation presented considered that the plasma species follow Maxwellian distribution functions of momenta, which are not affected; only the average densities will change in time. This model can be improved considering that the absorption of plasma particles will depend on their momenta, so that the initial distribution function may be deviated from the Maxwellian form. To do this, it is necessary to use the kinetic equations which, instead of describing the time evolution of plasma densities, describe the time evolution of the particle velocity distribution functions. We intend to follow this investigation considering a model similar to the model utilized in this work, and expect to publish the results in the near future.

\section*{Acknowledgements}

This study was financed in part by the Coordena{\c{c}}{\~a}o de Aperfei{\c{c}}oamento de Pessoal de N{\'i}vel Superior – Brasil (CAPES) – Finance Code 001. LFZ acknowledges support from CNPq (Brazil), grant No. 303189/2022-3. RG acknowledges support from CNPq (Brazil), grant No. 313330/2021-2. 

\section*{Author Declarations}
\subsection*{Conflict of Interest}
The authors have no conflicts to disclose.

\subsection*{Author Contributions}
\textbf{L. B. De Toni}: Conceptualization (equal); Investigation (equal); Writing/Original Draft Preparation (lead); Writing/Review \& Editing (equal). \textbf{L. F. Ziebell}: Conceptualization (equal); Investigation (equal); Writing/Review \& Editing (equal). \textbf{R. Gaelzer}: Conceptualization (equal); Investigation (equal); Writing/Review \& Editing (equal).

\section*{Data Availability}
The data that support the findings of this study are available from
the corresponding author upon reasonable request.

% Create the reference section using BibTeX:
\section*{References}

\bibliography{MAIN}

\end{document}